\documentclass[12pt]{article}
\usepackage{graphicx}

\newcommand\pubnumber{CIPANP2015-Mukhopadhyay}
\newcommand\pubdate{\today}

\def\napoli{Department of Physics, Indian Institute of Science,\\
Bangalore 560012, INDIA}

\def\Title#1{\begin{center} {\Large #1 } \end{center}}
\def\Author#1{\begin{center}{ \sc #1} \end{center}}
\def\Address#1{\begin{center}{ \it #1} \end{center}}

\newcommand\pubblock{\rightline{\begin{tabular}{l} \pubnumber\\
         \pubdate  \end{tabular}}}
\newenvironment{Abstract}{\begin{quotation}  }{\end{quotation}}
\newenvironment{Presented}{\begin{quotation} \begin{center} 
             PRESENTED AT\end{center}\bigskip 
      \begin{center}\begin{large}}{\end{large}\end{center} \end{quotation}}
\def\Acknowledgements{\bigskip  \bigskip \begin{center} \begin{large}
             \bf ACKNOWLEDGEMENTS \end{large}\end{center}}
\def\beq{\begin{equation}}
\def\eeq#1{\label{#1}\end{equation}}
\def\eeqn{\end{equation}}

\def\beqa{\begin{eqnarray}}
\def\eeqa#1{\label{#1}\end{eqnarray}}
\def\eeqan{\end{eqnarray}}

\let\bar=\overbar

\def\Dslash{\not{\hbox{\kern-4pt $D$}}}
\def\dslash{\not{\hbox{\kern-2pt $\del$}}}

\def\msb{{\bar{\ssstyle M \kern -1pt S}}}

\begin{document}
\begin{titlepage}
\pubblock

\vfill
\Title{Significantly super-Chandrasekhar limiting mass white dwarfs as progenitors for
peculiar over-luminous type Ia supernovae}
\vfill
\Author{ Banibrata Mukhopadhyay}
\Address{\napoli}
\vfill
\begin{Abstract}
Since 2012, we have initiated developing systematically the simplistic to rigorous models to prove that
highly super-Chandrasekhar, as well as highly sub-Chandrasekhar, limiting mass
white dwarfs are possible to exist.
We show that the mass of highly magnetized or modified Einstein's gravity induced white dwarfs could be significantly 
super-Chandrasekhar and such white dwarfs could altogether have a different mass-limit. On the other hand, type Ia 
supernovae (SNeIa), a key to unravel the evolutionary history of the universe, are believed to be triggered in white 
dwarfs having mass close to the Chandrasekhar-limit. However, observations of several peculiar, over- and under-luminous 
SNeIa argue for exploding masses widely different from this limit. We argue that explosions of super-Chandrasekhar 
limiting mass white dwarfs result in over-luminous SNeIa. 
We arrive at this revelation, first by considering simplistic, spherical, Newtonian 
white dwarfs with constant magnetic fields. Then we relax the Newtonian assumption and consider
the varying fields, however obtain similar results. Finally, we consider a full scale general 
relativistic magnetohydrodynamic description of white dwarfs allowing their self-consistent departure 
from a sphere to ellipsoid. Subsequently, we also explore the effects of modified Einstein's gravity.
Our finding questions the uniqueness of the Chandrasekhar-limit. 
It further argues for a possible second standard candle, which has many far reaching implications.
\end{Abstract}
\vfill
\begin{Presented}
CIPANP2015, Vail, CO, USA, May 19--24, 2015
\end{Presented}
\vfill
\end{titlepage}
\def\thefootnote{\fnsymbol{footnote}}
\setcounter{footnote}{0}

\section{Introduction}

Since last $2-3$ years, we have been exploring physics behind the origin of peculiar, over-luminous type 
Ia supernovae (SNeIa)
by invoking super-Chandrasekhar white dwarfs as their progenitor.
Since our proposal, idea of super-Chandrasekhar white dwarfs has come into lime-light ---
so many follow-up papers have appeared subsequently. 
It was indeed argued earlier, based on observation, that such peculiar, over-luminous SNeIa would exhibit super-Chandrasekhar 
progenitors \cite{howel}. 

The peculiarity of over-luminous SNeIa not only lies with their power, but also with the shape
of their lightcurves. All SNeIa generally follow a luminosity-stretch relation --- Philip's relation \cite{phil} ---
larger the peak luminosity, slower is the declining rate of the power and vice versa. Such a relation
does not hold in peculiar SNeIa --- the larger peak luminosity corresponds to the faster declining rate of power.
Interestingly, the kinetic energy of these SNIa ejecta is observed to be very low, hence higher power
in the lightcurve could only be explained by invoking a larger progenitor mass.

Any new idea, when proposed, generally is tested with a simplistic model first. Once, the results 
based on a simplistic model show promise to explain observations and/or experiments, then more realistic
self-consistent models, introducing more sophisticated physics, are introduced in order to 
fine-tune the original model. Without being an exception, we have also followed the same tactics
to develop our super-Chandrasekhar white dwarf model.

We have, so far, approached towards this mission through the following steps. First, we have 
considered most simplistic, spherically symmetric, highly magnetized white dwarfs in the Newtonian 
framework, assuming the magnetic field to be constant or almost constant throughout (or modeling,
as if, the inner region of white dwarfs) \cite{prd}. However, it has been speculated in that work itself
that the self-consistent consideration of deformation of white dwarfs would reveal a similar
super-Chandrasekhar mass at lower fields. In the same model framework, we have also shown that
magnetized white dwarfs altogether have a new mass-limit, $80\%$ larger than the Chandrasekhar-limit \cite{prl},
in the same spirit as the Chandrasekhar-limit was obtained \cite{chandra}.
Afterwards, we have removed both the assumptions: the Newtonian description and spherical symmetry
(e.g. \cite{jcap14,jcap15a}).
Note that magnetized white dwarfs could be significantly smaller in size compared to their
conventional counter-parts \cite{prd,prl} and, hence, general relativistic effects therein may not be negligible.
Thus, based on a full scale general relativistic magnetohydrodynamic (GRMHD) description \cite{pili},
we have explored more self-consistent white dwarfs which are ellipsoid and have revealed
similar stable masses as obtained in the simpler framework \cite{jcap15a}. 

All the above explorations are, however, based on the power of magnetic fields. We have also explored the effects
of a possible modification to Einstein's gravity. Note that apart from over-luminous ones, some of
SNeIa are under-luminous as well. By invoking the modified Einstein's gravity (in the first instance,
simplistic Starobinsky gravity), we have shown that, depending on their density, white dwarfs
could have significantly super- as well as sub-Chandrasekhar limiting masses. This apparently
unifies two apparently disjoint classes of SNeIa and argues that Einstein's gravity theory may
not be the ultimate theory even in the stellar physics. 
 
In the next sections, we describe all the explorations one by one to firmly establish our theory. Our attempt will be to 
demonstrate significantly super- and sub-Chandrasekhar limiting masses, rather than 
their explosions to give rise to the peculiar, respectively, over- and under-luminous 
SNeIa. We will assume that such white dwarfs, on approaching their respective 
super- and sub-Chandrasekhar limiting masses, will reveal over- and under-luminous
SNeIa respectively, whose proof will be deferred for a future work.

\section{Model I: Spherical white dwarfs with constant magnetic fields in the Newtonian framework}

In the presence of a strong uniform magnetic field, 
the energy states of a free electron are quantized into Landau orbitals.
%which define the motion of
%the electron in a plane perpendicular to the magnetic field. 
Theoretically, the 
Landau quantization effects start affecting electrons at a field $B_c=4.414\times 10^{13}$G and above, although practically  
it requires at least another order of magnitude higher field to affect them.

The Fermi energy level ($E_F$) of a Landau quantized electron is given by
\begin{eqnarray}
E_{F}^{2} = p_{F}(\nu)^{2}c^{2} + m_{e}^{2}c^{4}\left(1 + 2\nu B_{D}\right),
\end{eqnarray}
when $p_F(\nu)$ is the Fermi momentum of $\nu$th Landau level ($\nu=0,1,2.....$), $m_e$ the 
mass of electrons, $c$ the speed of light, $B_D$
the magnetic field in units of $B_c$. 
%From the condition that $p_{F}(\nu)^{2} \geq 0$, the maximum number of occupied Landau
%levels is given by \cite{lai}
%\begin{equation}
%\nu_m= \frac{\left(\frac{E_{Fmax}}{m_ec^2}\right)^{2} - 1}{2B_{D}}. 
%\label{be}
%\end{equation}
The Fermi energy of electrons in units of $m_{e}c^{2}$ for a given $\nu$ is given by
\begin{equation}
\epsilon_{F}^{2} = x_{F}(\nu)^{2} + (1 + 2\nu B_{D}),
\label{fermi}
\end{equation}
where $x_{F}(\nu)$ is the Fermi momentum in units $m_{e}c$.
As $x_{F}(\nu)^{2} \geq 0$, the maximum number of occupied Landau levels
\begin{equation}
\nu_{m} = \left(\frac{\epsilon_{Fmax}^{2} - 1}{2B_{D}}\right)_{\rm nearest ~ lowest ~ integer}.
\label{max}
\end{equation}
For an one Landau level system, when only ground Landau level ($\nu=0$) is occupied, $\nu_m=1$.
Similarly, for a two level system, when ground ($\nu=0$) and first ($\nu=1$) levels are occupied, 
$\nu_m=2$, and so on.

We write \cite{lai} the electron number density
\begin{equation}
n_{e} = \frac{2B_{D}}{(2\pi)^{2} \lambda_{e}^{3}} \sum_{\nu=0}^{\nu_{m}} g_{\nu}x_{F}(\nu),
\label{ne}
\end{equation}
where the Compton wavelength of the electron $\lambda_{e} = \hbar/m_{e}c$ and 
$g_\nu$ is the degeneracy that arises due to the Landau level splitting, such that,
$g_\nu = 1$ for $\nu = 0$ and $g_\nu = 2$ for $\nu\ge 1$, the matter density
\begin{equation}
\rho = \mu_{e}m_{H}n_{e},
\label{mat}
\end{equation}
where $\mu_{e}$ is the mean molecular weight per electron and $m_{H}$ the mass of hydrogen atom, 
the electron degeneracy pressure
\begin{eqnarray}
P =  \frac{2B_{D}}{(2\pi)^{2}\lambda_{e}^{3}}m_{e}c^{2}\sum_{\nu=0}^{\nu_{m}} g_{\nu} (1 + 2\nu B_{D}) \eta\left(\frac{x_{F}(\nu)}{(1 + 2\nu B_{D})^
{1/2}} \right), 
\label{pressure}
\end{eqnarray}
where
\begin{equation}
{\eta}(y) = \frac{1}{2}y\sqrt{1 + y^{2}} - \frac{1}{2}\ln(y+ \sqrt{1 + y^{2}}).
\end{equation}

%Hence, for an one Landau level system, 
%density ($\rho$) and pressure ($P$) 
%of the electron degenerate gas \cite{km} are given by
%\begin{eqnarray}
%\rho=\frac{\mu_em_H B_D}{2\pi^2\lambda_e^3}\sqrt{\left(\frac{E_F}{m_ec^2}\right)^2-1}\longrightarrow 
%\frac{\mu_em_H }{2\pi^2m_e c^2\lambda_e^3}B_DE_F\,\,\,{\rm for}\,\,\,E_F>>m_ec^2,
%\label{rho}
%\end{eqnarray}
%\begin{eqnarray}
%P=\frac{B_Dm_ec^2}{4\pi^2\lambda_e^3}\left(\frac{E_F\sqrt{E_F^2-m_e^2c^4}}{(m_ec^2)^2}
%-\ln\left\{\frac{E_F+\sqrt{E_F^2-m_e^2c^4}}{m_ec^2}\right\}\right),
%\label{pres}
%\end{eqnarray}
%when $\mu_e$ is the mean molecular weight, $m_H$ the mass of proton, 
%$\lambda_e$ the Compton wavelength of electron.
%Now eliminating $E_F$ from equations (\ref{rho}) and (\ref{pres}) for an arbitrary $E_F$, we arrive at 
%the equation of state (EoS) is given by
%\begin{eqnarray}
%P=\frac{m_ec^2}{2Q\mu_em_H}\left(\rho\sqrt{Q^2+\rho^2}-Q^2\ln\left\{\frac{\rho+\sqrt{Q^2+\rho^2}}
%{Q}\right\}\right),
%\label{eos}
%\end{eqnarray}
%when $Q=\mu_e m_H B_D/2\pi^2\lambda_e^3$.

%Following previous work \cite{dasmu}, we now approximate the above EoS and EoSs for any other $\nu_m$
%by a polytropic relation 
%$P=K\rho^\Gamma$ such that the polytropic
%index $\Gamma=1+1/n$ is piecewise constant in different density ranges  and $K$ being a dimensional 
%constant. This will prove to be useful in order to understand 
%scaling behaviors of mass and radius of the white dwarf with its central density.

%Now in terms of $Q$, equation (\ref{rho}) can be rewritten as
%\begin{eqnarray}
%\rho=Q \frac{E_F}{m_e c^2} .
%\label{Qrho}
%\end{eqnarray}
In the limit $E_F>>m_ec^2$, which corresponds to a very high density (as well as a high magnetic field and, hence, $\nu_m=1$), 
%$\rho >> Q$ \cite{prd,apjl}.  Thus 
combining equations (\ref{ne}), (\ref{mat}) and (\ref{pressure}) we obtain
%\begin{equation}
%P=\frac{m_ec^2}{2Q\mu_em_H}\left(\rho^2-Q^2\ln\left\{\frac{2\rho}{Q}\right\}\right) . 
%\end{equation}
%The logarithmic term is much smaller than the first term in the above equation and hence by
%neglecting it we obtain
\begin{equation}
P=\frac{m_ec^2}{2Q\mu_em_H}\rho^2=K_m\rho^\Gamma,
\label{km}
\end{equation}
which corresponds to the polytropic EoS with $\Gamma=2$.

The underlying magnetized, spherical white dwarf obeys the magnetostatic equilibrium 
condition 
\begin{eqnarray}
\frac{1}{\rho+\rho_B}\frac{d}{dr}\left(P+\frac{B^2}{8\pi}\right)=F_g+\left.\frac{\vec{B}\cdot\nabla\vec{B}}{4\pi(\rho+\rho_B)}\right|_r,
\label{magstat}
\end{eqnarray}
when $r$ is the radial distance from the center of white dwarf, $\vec{B}$ the magnetic field 
in G, $B^2=\vec{B}\cdot\vec{B}$, $F_g$ the radial component of gravitational force and $\rho_B$ the 
magnetic density. This equation is supplemented
by the equation for the estimate of mass ($M$) within any $r$, given by
\begin{eqnarray}
\frac{dM}{dr}=4\pi r^2(\rho+\rho_B).
\label{mas}
\end{eqnarray}
The magnetic field in the white dwarf is assumed to be very slowly varying such that the combined effect of 
magnetic pressure gradient and tension is cancelled out by the effect of gravity due to the magnetic density
(as justified previously \cite{dasmu2,dasmu}). 
Moreover, at a very large density (in the limiting case considered here in the spirit of the Chandrasekhar-limit), 
the star becomes so compact as if the magnetic field remains (almost) constant throughout.
Hence, taking above facts into consideration (which effectively converts the magnetostatic balance condition to hydrostatic 
condition) and combining equations (\ref{magstat}) and (\ref{mas}),
we obtain
\begin{equation}
\frac{1}{r^{2}}\frac{d}{dr}\left(\frac{r^{2}}{\rho}\frac{dP}{dr} \right) = -4\pi G \rho,
\label{diff}
\end{equation}
where $G$ is the Newton's gravitation constant.
%We now briefly recall the Lane-Emden formalism (see, e.g., \cite{arc}). 
Let us define
\begin{equation}
\rho = \rho_{c}\theta^{n},~~~~r = a\xi,
\label{theta}
\end{equation}
where $\rho_{c}$ is the central density of the white dwarf, $\theta$ a dimensionless variable,
$n=1/(\Gamma-1)$,
%and
%\begin{equation}
%r = a\xi,
%\label{xi}
%\end{equation}
$\xi$ another dimensionless variable and constant $a$ carries the dimension of 
length defined as
\begin{equation}
a = \left [\frac{(n+1)K_m\rho_{c}^{\frac{1-n}{n}}}{4\pi G} \right ]^{1/2}.
\label{a}
\end{equation}
Thus using equations (\ref{km}), (\ref{theta}) and (\ref{a}), 
equation (\ref{diff}) reduces to 
\begin{equation}
\frac{1}{\xi^{2}}\frac{d}{d\xi}\left(\xi^{2} \frac{d\theta}{d\xi} \right) = - \theta^{n},
\label{lane}
\end{equation}
which is the famous Lane-Emden equation.
Equation (\ref{lane}) can be solved for a given $n$, along with the boundary conditions
\begin{equation}
\theta(\xi = 0) = 1,~~~~\left (\frac{d\theta}{d\xi} \right)_{\xi=0} = 0.
\label{bc1}
\end{equation}
%and
%\begin{equation}
%\left (\frac{d\theta}{d\xi} \right)_{\xi=0} = 0.
%\label{bc2}
%\end{equation}
Note that 
%for $n < 5$, $\theta$ becomes zero for a finite value of $\xi$, say $\xi_{1}$, 
the surface of white dwarf corresponds to $\xi=\xi_1$ when $\theta=0$,
%which basically corresponds to the surface of the white dwarf 
such that its radius 
\begin{equation}
R = a\xi_{1}.
\label{R}
\end{equation}
Also by combining equations (\ref{mas}), (\ref{theta}) and (\ref{lane}), we obtain the 
mass of the white dwarf
\begin{equation}
M = 4\pi a^{3} \rho_{c}\int \limits_{0}^{\xi_{1}} \xi^{2}\theta^{n}\, d\xi.
\label{M}
\end{equation}

Now, the scalings of mass and radius
of the white dwarf with $\rho_c$ are easily obtained as
\begin{eqnarray}
M \propto K_m^{3/2} \rho_c^{(3-n)/2n},\,\,\,\, R\propto K_m^{1/2}\rho_c^{(1-n)/2n}. 
\label{scal}
\end{eqnarray}
Clearly $n=3$ ($\Gamma=4/3$) corresponds to $M$ independent of $\rho_c$, provided $K_m$ is
independent of $\rho_c$,
and hence limiting mass. 
%Therefore, we have to find out the condition for which $n=3$ and the corresponding
%proportionality constant for the scaling of $M$.
%Note, however, that $n=1$ for the extremely magnetized, highly dense, degenerate electron gas EoS,
%which is the present regime of interest.

However, in the extreme condition of magnetized white dwarfs, from equations (\ref{max}) and (\ref{km}), we obtain 
\begin{equation}
K_m=K\rho_c^{-2/3},
%K_m=\frac{c\hbar\pi^{2/3}}{2^{1/3}(m_H\mu_e)^{4/3}}\rho_c^{-2/3},
\end{equation}
where $K$ is a constant
and hence from equation (\ref{scal})
\begin{equation}
M\propto\rho_c^{3(1-n)/2n},\,\,\,R\propto\rho_c^{(3-5n)/6n},
\label{massf}
\end{equation}
revealing $M$ independent of $\rho_c$ for $n=1$, when
the radius becomes independent of the mass in the mass-radius
relation \cite{apjl,prl}.
%\begin{equation}
%K_m=K\rho_c^{-2/3},
%K_m=\frac{c\hbar\pi^{2/3}}{2^{1/3}(m_H\mu_e)^{4/3}}\rho_c^{-2/3},
%\end{equation}
Now combining equations
(\ref{a}), (\ref{M}) with $n=1$, we obtain the value of limiting mass
\begin{equation}
M_{l}=\frac{5.564}{\mu_e^2}\left(\frac{c\hbar}{G m_H^{4/3}}\right)^{3/2},
\label{mass3}
\end{equation}
when the limiting radius $R_l\rightarrow 0$.
For $\mu_e=2$, $M_l=2.58M_\odot$, where $M_\odot$ is solar mass. Importantly, 
for finite but high density and magnetic field, e.g. $\rho_c=2\times 10^{10}$gm/cc
and $B=8.8\times 10^{15}$G when $E_{Fmax}=20m_ec^2$, $M=2.44M_\odot$ and $R$ is about 650km. Note that
above $\rho_c$ and $B$ are below their respective upper limits set by the instabilities of
pycnonuclear fusion, inverse-$\beta$ decay and general relativistic effects \cite{dasmu}, however
still produce a significantly super-Chandrasekhar smaller white dwarf with a mass very close to the limiting value.

\section{Model II: Spherical white dwarfs with varying magnetic fields in the general relativistic framework} 

It is important to note that
a large number of white dwarfs have been discovered by the Sloan Digital Sky Survey (SDSS),
having high surface fields $10^5-10^9$G \cite{schmidt03,vanlandingham05}.
It is likely that the observed surface field is several orders of magnitude smaller
than the central field. Thus, it is important to perform investigations, in the presence
of varying fields, by accounting $dP_B/dr$ and $\rho_B$, in addition to $dP/dr$.
Keeping this in mind, we model the variation of the magnitude of magnetic field as a function of $\rho$ \cite{bando}
given by
\begin{equation}
B \left(\frac{\rho}{\rho_0}\right) = B_s + B_0\left[1-\exp \left(-\eta \left(\frac{\rho}{\rho_0}\right)^\gamma \right)\right],
\label{bprofile}
\end{equation}
%where $\rho_0$ is a density near the surface of the corresponding white dwarf 
where $\rho_0$, for the present purpose, is chosen to be one-tenth of $\rho_c$
of the corresponding white dwarf,
$B_s$ the surface magnetic field, $B_0$ a parameter having
dimension of $B$, $\eta$ and $\gamma$ are dimensionless parameters which determine how exactly the
field magnitude decays from the center to the surface. 
%Note that as $\rho \rightarrow 0$ close to the surface of the white dwarf, $B \rightarrow B_s$.
We consider the cases with $10^9$G $\leq B_s \leq 10^{12}$G. 
%for $B_s$ is guided by the aforementioned SDSS observations.
However, for the central magnetic field $B_{\rm cent}\geq 10^{14}$G, the result is independent of the above considered
$B_s$ or less.

In the general relativistic framework, the spherically symmetric white dwarfs are described by
the (magnetized) Tolman-Oppenheimer-Volkoff (TOV) equations \cite{jcap14,wald}, given by
\begin{equation}
\frac{dM(r)}{dr} = 4\pi r^2 (\rho(r)+\rho_B),
\label{tov_mass}
\end{equation}
\begin{equation}
\frac{d\rho(r)}{dr} = -\frac{G\left(\rho(r)+\rho_B + \frac{(P(r)+P_B)}{c^2}\right)\left(M(r)+ \frac{4\pi r^3 (P(r)+P_B)}{c^2}\right)}{r^2\left(1-\frac{2GM(r)}{c^2r}\right)\left(\frac{dP(r)}{d\rho} +\frac{dP_B}{d\rho}\right)},
\label{tov_dens}
\end{equation}
assuming the averaged
effects of magnetic field to the pressure and hence the star is isotropic such that the 
magnetic pressure $P_B=B^2/24\pi$.
Validity of such a consideration has been already justified earlier
\cite{dasmu,adam,cheoun,herrera}.

\begin{figure}[htb]
\centering
\includegraphics[height=3.0in]{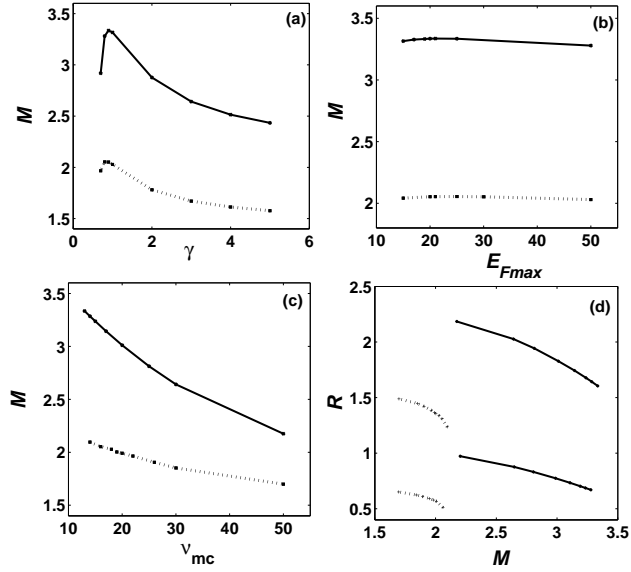}
\caption{
Super-Chandrasekhar white dwarfs with varying magnetic fields --- the solid and dotted lines
represent the cases corresponding to constraints (i) and (ii) respectively (see text).
(a) $M$ as a function of $\gamma$ for $E_{Fmax}=20m_ec^2$, $B_{\rm cent}=6.77\times 10^{14}$G,
maximum $\eta$, upto which constraints (i) and (ii) satisfy, for respective $\gamma$s. (b) $M$ as 
a function of $E_{Fmax}$ for $B_{\rm cent}=6.77\times 10^{14}$G,
$\gamma=0.9$ for solid line and $B_{\rm cent}=5.18\times 10^{14}$ G, $\gamma=0.8$ for dotted line,
for the respective best $\eta$s. (c) $M$ as a function
of $\nu_{mc}$ for $E_{Fmax}=20m_ec^2$, $\gamma=0.9$ for solid line and $\gamma=0.8$ for dotted line,
for the respective best $\eta$s.
(d) The topmost solid and dotted lines represent the $M$-$R$ relations corresponding to (c), while the solid and
dotted lines at the bottom represent the $M$-$R$ relations corresponding to (c) but with $E_{Fmax}=50m_ec^2$.
$M$, $R$ and $E_{Fmax}$ are in units of $M_\odot$, 1000km and $m_ec^2$ respectively.
}
\label{mass}
\end{figure}

In the presence of a strong magnetic field, the total pressure of the system
in the direction perpendicular to the field is $P_{\perp}=P+B^2/8\pi$ (neglecting magnetization, which is much smaller compared
to $B^2/8\pi$ for the field of present interest), while that in the direction parallel to
the field is given by $P_{||}=P-B^2/8\pi$. Hence, we propose two plausible constraints on
the magnetic field profile in order to avoid any instability due to negative parallel pressure, 
such that: (i) the average parallel
pressure, given by $P-P_B$, should remain positive throughout the white dwarf, (ii) $P_{||}$ should remain positive 
throughout the white dwarf (see \cite{jcap14} for details).
Various results obtained by considering different magnetic field profiles, 
the number of occupied Landau levels at the center $\nu_{mc}$, and $E_{Fmax}$,
are summarized in Figure \ref{mass}. 
We restrict $E_{Fmax}$ to $50 m_e c^2$, in order to avoid possible neutronization of the matter.
%\begin{figure*}
%\begin{center}
%\includegraphics[angle=0,width=18cm]{fig_4block_mr_20n50big.eps}
%\caption{ Super-Chandrasekhar white dwarfs with varying $B$ --- The solid and dotted lines
%represent the cases corresponding to constraints (i) and (ii) respectively (see text).
%(a) $M$ as a function of $\gamma$ for $E_{Fmax}=20m_ec^2$, $B_{\rm cent}=6.77\times 10^{14}$ G,
%$\eta=\eta_{max}$ for respective $\gamma$s. (b) $M$ as a function of $E_{Fmax}$ for $B_{\rm cent}=6.77\times 10^{14}$ G,
%$\gamma=0.9$ for solid line and $B_{\rm cent}=5.18\times 10^{14}$ G, $\gamma=0.8$ for dotted line. (c) $M$ as a function
%of $\nu_{mc}$ for $E_{Fmax}=20m_ec^2$, $\gamma=0.9$ for solid line and $\gamma=0.8$ for dotted line.
%(d) The topmost solid and dotted lines represent the $M$-$R$ relations corresponding to (c), while the solid and
%dotted lines at the bottom represent the $M$-$R$ relations corresponding to (c) but with $E_{Fmax}=50m_ec^2$.
%$M$, $R$ and $E_{Fmax}$ are in units of $M_\odot$, 1000 km and $m_ec^2$ respectively.
%  }
%\label{mass}
%\end{center}
%\end{figure*}
%\begin{figure}[t]

Figure \ref{mass}(d) depicts the most important results, i.e. the mass-radius relations.
Note that the radii of the
white dwarfs having $E_{Fmax}=50m_ec^2$ are more than a factor of two smaller than those with $E_{Fmax}=20m_ec^2$ for
roughly the same range of $M$. This is expected because a higher $E_{Fmax}$ implies a
higher $\rho_c$ and hence more compact objects. For example, for the cases pertaining to the constraint (i),
$M_{\rm max}=3.33M_\odot$ and $R=1605$km for $E_{Fmax}=20m_ec^2$ ($\rho_c=1.55\times 10^{10}$gm/cc),
while $M_{\rm max}=3.28 M_\odot$ and $R=670$km for $E_{Fmax}=50m_ec^2$ ($\rho_c=2.42\times 10^{11}$gm/cc).
Similarly, for the cases pertaining to the constraint (ii), $M_{\rm max}=2.1M_\odot$ and $R=1237$km for $E_{Fmax}=20m_ec^2$,
while $M_{\rm max}=2.06 M_\odot$ and $R=512$km for $E_{Fmax}=50m_ec^2$.

\section{Model III: Most self-consistent spheroidal white dwarfs with varying magnetic fields 
in the general relativistic framework}

Now we explore, using the (modified) {\it XNS} code \cite{pili,xns1},
GRMHD analyses of magnetized, rotating white dwarfs and confirm the validity of all the preceding results.
%Note that the {\it XNS} code is particularly suited to solving for highly deformed stars due to a strong
%magnetic field and/or rotational effects (\cite{xns1,pili}).
Although, originally the {\it XNS} code was developed to investigate deformed neutron stars, recently it was
modified \cite{jcap15a,sathya} in order to obtain
equilibrium configurations of deformed, magnetized, rotating white dwarfs.
We self-consistently find that for a range $10^{10} \le \rho_c \le 10^{11}$gm/$\rm cm^3$, 
the maximum magnetic field strength inside the white dwarf ranges as $10^{13} \le B_{\rm max} \le 10^{15}$G. 
Consequently, $\nu_m \ge 20$ for this range of $\rho_c$ and $B_{\rm max}$, which does not 
significantly modify $\Gamma$ (see, e.g., \cite{prd}). 
We consider for all the computations $\rho_c = 1.9902\times 10^{10}$gm/cc, which is high enough to ensure almost 
complete relativistic electron degeneracy, so that we may use the polytropic EoS with $n=3$ (i.e. $\Gamma={4}/{3}$) 
consistently throughout the star. This density is still lower than the limit at which gravitational instabilities 
in general relativity set in, which is about $3\times 10^{10}$gm/cc.

We refer the readers to the previous works \cite{pili,jcap15a,xns1,sathya}
for a complete description of GRMHD formulation, which includes the equations characterizing the geometry 
of the magnetic field and the underlying current distribution, as well as the numerical technique 
employed by the {\it XNS} code to solve them. 
As mentioned earlier, the presence of a strong magnetic field in a compact object generates an anisotropy in the 
magnetic pressure which in turn causes the star to be deformed, which we consider here self-consistently.
Of course, the degree of anisotropy depends on the strength and geometry of the magnetic field.
We construct axisymmetric white dwarfs in spherical polar coordinates $(r,\theta,\phi)$,
self-consistently accounting for the deviation from spherical symmetry.
Stationary configurations can have many rotation laws consistent with the magnetostatic balance condition.

\begin{figure}[htb]
\centering
\includegraphics[height=3.5in]{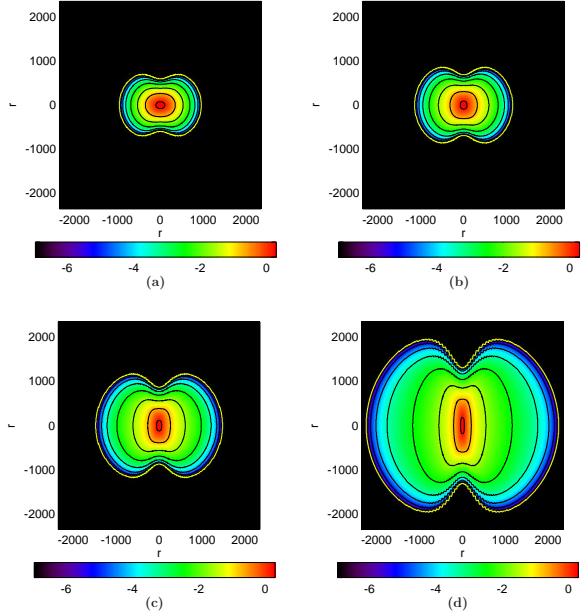}
\caption{
Sequence of differentially rotating white dwarfs with a purely toroidal magnetic
field, with changing $B_{\rm max}$ and fixed $\Omega_c=30.42$sec$^{-1}$.
The panels are contour plots of
log$\left(\frac{\rho}{\rho_0}\right)$, with $\rho_0=10^{10}$gm/cc, corresponding to the $B_{\rm max}/10^{14}{\rm G}$ values (a) 0,
(b) $2.299$, (c) $2.996$, (d) $3.584$.
The corresponding physical quantities are listed in Table 1. 
}
\label{self-mass}
\end{figure}

Figure \ref{self-mass} illustrates a typical set of GRMHD solutions for magnetized, differentially rotating white dwarfs.
With the increase of toroidally dominated magnetic field, upto the maximum value considered $B_{\rm max}=3.584\times 10^{14}$G
(do not confuse with $B_{\rm max}$ in a given white dwarf defined in Model II), the 
mass increases up to $3.159M_\odot$ --- by nearly $78\%$
from the mass in the non-magnetised case. 
The equatorial radius $r_e$ for $B_{\rm max}$ becomes about 3322km, more than double that of the non-magnetized configuration.
The surface angular velocity $\Omega_{eq}$ reduces from 2.99 to 0.593sec$^{-1}$.
This could be understood as the increase of $r_e$ increases
the centrifugal force at the equator for a given $\Omega_{eq}$, which enforces decreased $\Omega_{eq}$ in order
to obtain equilibrium solutions. See Table~1 for details.
We furthermore note the polar concavities developing primarily due to the
differential rotation, are accentuated by the toroidal field. 
%Figure \ref{self-mass} clearly shows the severe polar denting with the increase in $B_{max}$.  
%See Table~1 for other related parameters and solutions.
Note that chosen toroidally dominated fields here have already been shown to be stable
and realistic based on the prescription for twisted-torus geometries \cite{cio}. However, the
poloidally dominated magnetized stable white dwarfs have also been shown to be possible with much smaller
radii \cite{sathya}, $<1000$km.

% float containing table and contour plots
% configs given correspond to XNS bcoef = 0, 35, 74, 150
%\begin{table}
%\centering
%\resizebox{\linewidth}{!}{
%\begin{tabular}{|c|c|c|c|c|c|c|}
%\hline
%$B_{max}$  & $M$ & $r_e$ & $\Omega_{eq}$ & KE/GE & ME/GE & $r_p/r_e$\tabularnewline
%\hline
%\hline
%0 & 1.769 & 1410 & 2.990 & 0.126 & 0 & 0.613 \tabularnewline
%\hline
%2.299 & 1.959 & 1676 & 2.180 & 0.132 & 0.046 & 0.603 \tabularnewline
%\hline
%2.996 & 2.318 & 2171 & 1.339 & 0.136 & 0.108 & 0.583 \tabularnewline
%\hline
%3.584 & 3.159 & 3322 & 0.593 & 0.132 & 0.203 & 0.584 \tabularnewline
%\hline
%\end{tabular}}
%\caption{Differentially rotating configurations with purely toroidal magnetic
%field, with changing $B_{max}$ in units of $10^{14}$G and $\Omega_c=30.42$~sec fixed.
%KE/GE and ME/GE are the ratios of kinetic and gravitational energies and magnetic to 
%gravitational energies respectively, $r_p$ is the polar radius}
%\end{table}

\begin{table}[t]
\begin{center}
%\begin{tabular}{l|ccc} 
\small
\begin{tabular}{|c|c|c|c|c|c|c|} \hline
$B_{\rm max} (10^{14}{\rm G})$  & $M(M_\odot)$ & $r_e({\rm km})$ & $\Omega_{eq}({\rm sec^{-1}})$ & KE/GE & ME/GE & $r_p/r_e$\\ \hline
0 & 1.769 & 1410 & 2.990 & 0.126 & 0 & 0.613 \\ 
2.299 & 1.959 & 1676 & 2.180 & 0.132 & 0.046 & 0.603 \\
2.996 & 2.318 & 2171 & 1.339 & 0.136 & 0.108 & 0.583 \\
3.584 & 3.159 & 3322 & 0.593 & 0.132 & 0.203 & 0.584 \\ \hline
%Patient &  Initial level($\mu$g/cc) &  w. Magnet &  
%w. Magnet and Sound \\ \hline
% Guglielmo B.  &   0.12     &     0.10      &     0.001  \\
% Ferrando di N. &  0.15     &     0.11      &  $< 0.0005$ \\ \hline
\end{tabular}
\caption{\small
Differentially rotating configurations with purely toroidal magnetic
field, with changing $B_{\rm max}$ and $\Omega_c=30.42$sec$^{-1}$ fixed.
KE/GE and ME/GE are the ratios of kinetic and gravitational energies and magnetic to
gravitational energies respectively, $r_p$ is the polar radius.
}
\label{tab:blood}
\end{center}
\end{table}

\section{Model IV: Spherical non-rotating, non-magnetized white dwarfs in the framework of
modified Einstein's gravity}

Let us consider the 4-dimensional action as 
\begin{equation}
S = \int \left[\frac{1}{16\pi} f(R) + {\cal L}_M \right] \sqrt{-g}~ d^4 x, 
\end{equation}
where $g$ is the determinant of the metric tensor $g_{\mu\nu}$,
%(which describes the nature of the underlying curvature of spacetime), 
${\cal L}_M$ the Lagrangian density of the matter field,
$R$ the scalar curvature defined as $R=g^{\mu\nu}R_{\mu\nu}$,
when $R_{\mu\nu}$ is the Ricci tensor and $f$ an
arbitrary function of $R$ (in Einstein's gravity $f(R)=R$). For the present purpose, we choose the Starobinsky model
\cite{starobi} defined as $f(R)=R+\alpha R^2$, when $\alpha$ is a constant such that $\alpha R<<1$, revealing
only the first order correction and $g_{\mu\nu}=g_{\mu\nu}^0+\alpha g_{\mu\nu}^1$, when $g_{\mu\nu}^0$ is the 
metric tensor in Einstein's gravity.
However, similar effects could also be obtained in other modified gravity theories, e.g.
Born-Infeld gravity (e.g. \cite{bana}). Now, on extremizing the above action,
one obtains the modified field equation as
\begin{equation}
G_{\mu\nu}+ \alpha \left[2 R G_{\mu\nu} + \frac{1}{2} R^2 g_{\mu\nu} - 2(\nabla_\mu \nabla_\nu - g_{\mu\nu}
\nabla_\mu \nabla^\mu)R \right] = 8\pi T_{\mu\nu},
\label{modfld}
\end{equation}
where $G_{\mu\nu}=R_{\mu\nu}-{g_{\mu\nu}}R/2$, is Einstein's field tensor, $T_{\mu\nu}$ the energy-momentum 
tensor of the matter field and $\nabla_\mu$ the covariant derivative.

Furthermore, we consider the hydrostatic equilibrium condition: $g_{\nu r}\nabla_\mu T^{\mu\nu}=0$,
with zero velocity. Hence, we obtain the {\it modified} TOV equations as
the differential equations for mass $M_\alpha (r)$, pressure $P_\alpha (r)$
(or density $\rho_\alpha (r)$)
and gravitational potential $\phi_\alpha (r)$, of spherically symmetric white dwarfs
(see \cite{jcap15b} for details).

The boundary conditions for the solutions of modified TOV equations 
are $M_\alpha (0)=0$ and $\rho_\alpha (0) =\rho_c$. 
The chosen EoS is same as that chosen for nonmagnetized 
white dwarfs \cite{chandra}, described in Model I, given by $P_0=K\rho_0^{1+(1/n)}$, for
extremely low and high densities, where $P$ and $\rho$ of Model I are replaced by 
$P_0$ and $\rho_0$ respectively in the spirit of linear perturbation.
%For low density, non-relativistic WDs, $n=3/2$ and $K=K_1$, 
%while for high density, relativistic WDs, $n=3$ and $K=K_2$.

\begin{figure}[htb]
\centering
\includegraphics[height=3.5in]{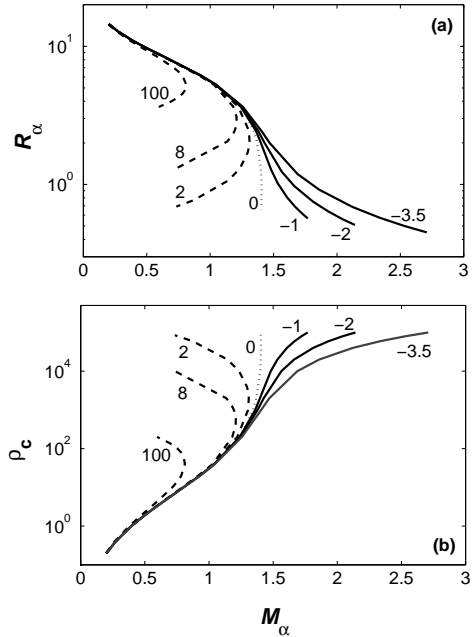}
\caption{
(a) Mass-radius relations. (b) Variation of $\rho_c$ with $M_\alpha$.
The numbers adjacent to the various lines denote
%the value of $\alpha_{13}$, which is defined as 
$\alpha/(10^{13}{\rm cm^2})$.
$\rho_c$, $M_\alpha$ and $R_\alpha$ are in units of $10^6$gm/cc, $M_\odot$ and 1000km respectively.
}
\label{mod}
\end{figure}

Figure \ref{mod} shows that
%that all three $M_\alpha-\rho_c$ curves for $\alpha<0$ overlap with the $\alpha=0$ curve in the low density region.
%However, 
with the increase of $\alpha$, the region of overlap of curves with the curve of Einstein's gravity ($\alpha=0$) 
recedes to a lower $\rho_c$. Modified Einstein's gravity
effects become important and visible at $\rho_c \ge 10^8,~4\times 10^7$ and $2\times 10^6$gm/cc,
for $\alpha = 2\times 10^{13}$, $8\times 10^{13}$ and
$10^{15}{\rm cm^2}$ respectively. For a given $\alpha$, with the increase of $\rho_c$, $M_\alpha$
first increases, reaches a maximum and then decreases, like the $\alpha=0$ case. With the
increase of $\alpha$, maximum mass $M_{\rm max}$ decreases and
for $\alpha=10^{15}{\rm cm}^2$ it is highly sub-Chandrasekhar ($0.81M_\odot$).
This feature of the modified gravity effect in white dwarfs was completely overlooked earlier. 
In fact, $M_{\rm max}$ for all the chosen $\alpha>0$ is sub-Chandrasekhar, ranging $1.31-0.81M_\odot$.
This is a remarkable finding since it establishes that even if $\rho_c$s for
these sub-Chandrasekhar white dwarfs are lower than the conventional value at which SNeIa are usually
triggered, an attempt
to increase the mass beyond $M_{\rm max}$ with increasing $\rho_c$ will lead to a gravitational instability.
This presumably will be followed by a
runaway thermonuclear reaction, provided the core temperature increases sufficiently due to collapse.
Occurrence of such thermonuclear runway reactions, triggered at densities as low as $10^6$gm/cc,
has already been demonstrated \cite{runaway}. Thus, once $M_{\rm max}$ is approached, a SNIa is expected to trigger just
like in the $\alpha=0$ case, explaining the under-luminous SNeIa \cite{1991bg,taub2008},
like SN 1991bg mentioned above.
%Table 2 confirms that the solutions for the $\alpha>0$ cases are within the perturbative regime. 

For $\alpha<0$ cases, Fig. \ref{mod} shows that for $\rho_c>10^8$ gm/cc, the
$M_\alpha-\rho_c$ curves deviate from the general relativity and
$M_{\rm max}$ for all the three cases corresponds to $\rho_c=10^{11}$gm/cc, which is  
an upper-limit chosen to avoid possible neutronization.
%that might occur for $\rho_c>10^{11}$ gm/cc. 
Interestingly, all values of $M_{\rm max}$ are highly super-Chandrasekhar, ranging
$1.8-2.7M_\odot$.
Thus, while the general relativity effect is very small, modified Einstein's gravity effect could
%, according to the perturbative $f(R)$-model, 
lead to $\sim 100\%$ increase in the limiting mass of white dwarfs, which was completely overlooked so far.
The corresponding values of $\rho_c$ are large enough to initiate thermonuclear reactions, e.g.
they are larger than $\rho_c$ corresponding to $M_{\rm max}$ of $\alpha=0$ case, whereas the respective
core temperatures are expected to be similar. This
explains the entire range of the observed over-luminous SNeIa mentioned above \cite{howel,scalzo}.

Hence, based on a single underlying theory, i.e. a modified Einstein's gravity, varying the 
value of $\alpha$, we obtain a range of sub- to super-Chandrasekhar limiting masses.
From Fig. \ref{mod}(b) it is clear that effectively $\alpha$ is a density dependent parameter,
and, hence, brings in the chameleon-like effect  (see, e.g., \cite{cham2}) in the model. In a more rigorous model, the 
quantity equivalent to $|\alpha\rho|$ could be an invariant quantity (see \cite{jcap15b} for details). 
Nevertheless, the main message here is that the modified gravity effect has a significant impact in white dwarfs.

\section{Summary and Conclusion}

We have shown, by developing systematically the simplistic to rigorous models,
that highly super-Chandrasekhar, as well as highly sub-Chandrasekhar,
limiting mass white dwarfs are possible to exist. While the effects of magnetic field 
render only highly super-Chandrasekhar mass-limit(s), the effects of modified Einstein's 
gravity reveal sub- and super- both the limits and, hence, apparently unify two disjoint classes
of SNeIa.

The new generic mass-limit of highly magnetized white dwarfs is in the range $2.6-3.4M_\odot$,
depending on the field profiles. Once the super-Chandrasekhar limiting mass is approached, the white dwarfs explode exhibiting
over-luminous, peculiar SNeIa. Indeed observations suggest the exploding mass to be in the range $2.3-2.8M_\odot$,
which tallies with our theoretical calculation.

Now SNeIa are used as a standard candle in order to understand the size,
and, hence expansion history of Universe, due the uniform mass of their progenitors. 
Now this `uniform mass' will no longer remain uniform if the progenitor mass-limit
is different --- possible second standard candle --- as we have established to be the case for
certain magnetized white dwarfs. If the peculiar SNeIa are eventually observed enormous in number, then
it might necessarily need to sample the observed data from supernova explosions
carefully which may affect the conclusion for expansion history of Universe.

Overall, our discovery raises two fundamental questions. Is the Chandrasekhar-limit unique?
Is Einstein's gravity the ultimate theory for understanding astronomical phenomena?
Both the answers appear to be no!

%%%%%%%%%%%%%%%%%%%%%%%%%%%%%%%%%%%%%%%%%%%%%%%%%%%%%%%%%%%%%%%%%%%%%%%%%
%%
%%   use this format to include an .eps figure into your paper
%%
%%%%%%%%%%%%%%%%%%%%%%%%%%%%%%%%%%%%%%%%%%%%%%%%%%%%%%%%%%%%%%%%%%%%%%%%%%%

%%%%%%%%%%%%%%%%%%%%%%%%%%%%%%%%%%%%%%%%%%%%%%%%%%%%%%%%%%%%%%%%%%%%%%%%%
%%
%%   use this format to include a LaTeX table  into your paper
%%
%%%%%%%%%%%%%%%%%%%%%%%%%%%%%%%%%%%%%%%%%%%%%%%%%%%%%%%%%%%%%%%%%%%%%%%%%%%

\Acknowledgements
I am thankful to Mukul Bhattacharya, Upasana Das and Sathyawageeswar Subramanian for
continuous discussions and working with me enthusiastically to evolve the topic. 
I am also thankful to Chanda J. Jog, Subroto Mukerjee and A. R. Rao for discussions and/or
collaborations towards this project.

\end{document}